\documentclass[twocolumn,showpacs,preprintnumbers,amsmath,amssymb,floatfix,nofootinbib]{revtex4-1}
\usepackage{amsmath,amsfonts,bbm,hyperref}
\usepackage{graphicx}% figures
\usepackage{epsfig}
\epsfclipon
\usepackage{enumitem} % list controls

\begin{document}

\preprint{UMN-TH-3804/18}

\title{Predicting the Sparticle Spectrum from Partially-Composite Supersymmetry}

\author{Yusuf Buyukdag, Tony Gherghetta, Andrew S. Miller}

\affiliation{%
\centerline{School of Physics and Astronomy, University of Minnesota, Minneapolis, Minnesota 55455, USA}
%\centerline{$^2$William I. Fine Theoretical Physics Institute, University of Minnesota, Minneapolis, MN 55455, USA}
Email: {\tt buyuk007@umn.edu, tgher@umn.edu, mill5738@umn.edu}}

\begin{abstract}
We use the idea of partial compositeness in a minimal supersymmetric model to relate
the fermion and sfermion masses. By assuming that the Higgs and third-generation matter is
(mostly) elementary, while the first- and second-generation matter is (mostly) composite, the
Yukawa coupling hierarchy can be explained by a linear mixing between elementary states and
composite operators with large anomalous dimensions. If the composite sector also breaks supersymmetry, 
then composite sfermions such as selectrons are predicted to be much heavier than the lighter elementary stops. This inverted sfermion mass hierarchy is consistent with current experimental limits that prefer light stops
($\mathcal{O}(10)$~TeV) to accommodate the 125~GeV Higgs boson, while predicting heavy first- and second-generation sfermions (${\gtrsim 100}$~TeV) as indicated by flavor physics experiments. The underlying dynamics can be modelled by a dual 5D gravity theory that also predicts a gravitino dark matter candidate ($\gtrsim$ keV), together with gauginos and Higgsinos, ranging from 10--90~TeV, that are split from the heavier first- and second-generation sfermion spectrum. This intricate connection between the fermion and sfermion mass spectrum can be tested at future experiments.
\end{abstract}

\maketitle

\section{Introduction}
\vspace{-4mm}

Supersymmetry provides a compelling theoretical framework for addressing some of the shortcomings of the Standard Model of particle physics. These include dark matter, gauge coupling unification, and the stabilization of the hierarchy between the electroweak and Planck scales. Because supersymmetry must be broken, the stability of the electroweak scale requires that the sparticle spectrum should not be too heavy. A vital clue for determining the superpartner mass scale comes from the recent discovery of the 125 GeV Higgs boson~\cite{Aad:2012tfa, Chatrchyan:2012xdj}. To obtain this mass in minimal supersymmetry, the Higgs quartic coupling must receive sizeable radiative corrections. These can arise from the top quark superpartners (or stops), provided that the lightest stop has mass of $\mathcal{O}(10)$~TeV. In the minimal framework there are no other sizeable contributions to the Higgs quartic coupling, and consequently the rest of the sparticle spectrum is not determined. The spectrum must only be compatible with the current LHC limits that require stop masses to be ${\gtrsim 1120}$~GeV~\cite{Sirunyan:2017xse} and gluino masses ${\gtrsim 1970}$~GeV~\cite{Aaboud:2017hrg}. Other indirect constraints such as the absence of flavor-changing processes, prefer the first- and second-generation scalar masses to be much heavier, ${\gtrsim 100}$~TeV. Thus, the current experimental situation seems to suggest a sizeable hierarchy in the \textit{sfermion} mass spectrum, which is inverted compared to the well-known \textit{fermion} mass hierarchy. For instance, the electron (top quark) is the lightest (heaviest) charged fermion, while the selectron (stop) may be the heaviest (lightest) charged sfermion.

In this Letter, we provide a mechanism that explains the origin of the inverted sfermion mass hierarchy and predicts the sparticle spectrum. The mechanism relies on partial compositeness~\cite{Kaplan:1991dc}, whereby the Standard Model fields are admixtures arising from the linear mixing of elementary states with composite operators. Assuming that the Higgs fields are elementary, the magnitudes of the corresponding Yukawa couplings then depend on the relative compositeness of the Standard Model fermions. To obtain an order-one Yukawa coupling with the Higgs, the top quark must be mostly elementary, while, since the elementary and composite sectors mix with an irrelevant coupling, the smallness of the electron Yukawa coupling follows from assuming that the electron is mostly composite. The remainder of the Standard Model Yukawa couplings are generated by varying degrees of compositeness. If one now further supposes that the composite sector is responsible for breaking supersymmetry, then an inverted hierarchy immediately follows. Selectrons, which are mostly composite, receive large supersymmetry-breaking masses, while stops, which are elementary, obtain suppressed supersymmetry-breaking contributions. In this way the fermion mass hierarchy determines the sfermion mass hierarchy and predicts an inverted mass spectrum.

The underlying strong dynamics that would be responsible for such a mechanism is similar to single-sector
models of supersymmetry breaking that were originally proposed in~\cite{ArkaniHamed:1997fq, Luty:1998vr},
with related work in~\cite{Gherghetta:2000kr, Gabella:2007cp, Franco:2009wf, Craig:2009hf, Aharony:2010ch}. Even if the underlying gauge theory were completely known, however, predictions for the spectrum would be difficult to obtain due to the nonperturbative dynamics. Therefore, we will instead use the AdS/CFT correspondence to model the strong dynamics with a
slice of AdS$_5$. In light of the Higgs boson discovery, this enables us to obtain specific quantitative predictions for the sparticle spectrum that can then be used to help guide future experimental searches.

\vspace{-4mm}
\section{Partially-Composite Supersymmetry}
\label{sec:PCsusy}
\vspace{-4mm}

To illustrate the mechanism of partially-composite supersymmetry, consider the elementary chiral superfield
$\Phi = \phi + \sqrt{2} \, \theta \psi + \theta \theta F$, where $\phi$ is a complex scalar, $\psi$ is a Weyl fermion,
and $F$ is an auxiliary field. In addition, we introduce a corresponding supersymmetric chiral operator
$\mathcal{O} = {\mathcal{O}_\phi} + \sqrt{2} \, \theta {\mathcal{O}_\psi} + \theta \theta{\mathcal{O}_F}$. The scaling dimensions of the component operators are $\operatorname{dim} \mathcal{O}_\phi = 1 + \delta_{\mathcal{O}} $, $\operatorname{dim} \mathcal{O}_\psi = \frac{3}{2} + \delta_{\mathcal{O}}$, and $\operatorname{dim} \mathcal{O}_F = 2 + \delta_{\mathcal{O}}$, where $\delta_{\mathcal{O}} \geq 0$ is the anomalous dimension~\cite{Cacciapaglia:2008bi}.

The supersymmetric Lagrangian contains separate elementary and composite sectors, together with linear mixing terms of the form $\left[ \Phi \, \mathcal{O}^c \right]_F$ for each chiral superfield $\Phi$ and charge-conjugate composite operator $\mathcal{O}^c$. At the UV scale $\Lambda_{\text{UV}}$, it is given by
\begin{equation}
  \mathcal{L}_\Phi
     = \left[ \Phi^{\dagger} \Phi \right]_D
    + \frac{1}{\Lambda_{\text{UV}}^{\delta-1}} \left( \left[ \Phi \, \mathcal{O}^c \right]_F + h.c. \right)~,
  \label{eq:chiralLag}
\end{equation}
where $\delta$ is the anomalous dimension of $\mathcal{O}^c$. We have taken order-one UV coefficients for the higher-dimension terms, and omitted a kinetic mixing between the elementary and composite sectors in our minimal setup. The composite sector is assumed to confine at an infrared scale, $\Lambda_{\text{IR}}$. In the limit of large-$N$ strong dynamics, the two-point function for the composite operator components can be written as a sum over one-particle states. In particular, for the scalar component, the two-point function $\langle \mathcal{O_\phi}(p) \mathcal{O_\phi}(-p) \rangle = \sum_{n} a_n^2/(p^2 + m_n^2)$
to leading order in $1/N$, where $a_n = \langle 0| \mathcal{O_\phi} |n \rangle \propto \sqrt{N}/(4 \pi)$ is the matrix element for $\mathcal{O_\phi}$ to create the $n^\text{th}$ state with mass $m_n$ from the vacuum~\cite{Witten:1979kh}.

The elementary-composite mixing in the Lagrangian \eqref{eq:chiralLag} mixes the elementary fields $(\phi, \psi)$ with the composite resonance states.  Assuming for simplicity just the lowest-lying composite state $\Phi^{(1)}=(\phi^{(1)},\psi^{(1)})$ with mass $g_\Phi^{(1)} \Lambda_{\text{IR}}$, the two-state system can be diagonalized to obtain the massless eigenstate~\cite{BGM}
\begin{equation}
   |\Phi_0 \rangle \simeq \mathcal{N}_\Phi
     \left\{| \Phi \rangle - \frac{1}{g_\Phi^{(1)} \sqrt{\zeta_\Phi}}
      \sqrt{\frac{\delta - 1}{\left( \frac{\Lambda_{\text{IR}}}{\Lambda_{\text{UV}}} \right)^{2 (1 - \delta)} - 1}} \,
       | \Phi^{(1)} \rangle\right\} \, ,
 \label{eq:scalaradmix}
 \end{equation}
where $\Phi_0 = (\phi_0,\psi_0)$, $g_\Phi^{(1)}$ and $\zeta_\Phi$ are order-one constants, and $\mathcal{N}_\Phi$ is a normalization constant. Given that $\Lambda_{\text{IR}}\ll \Lambda_{\text{UV}}$, this expression shows that the massless eigenstates are mostly elementary for $\delta > 1$, whereas for $0 \leq \delta < 1$ they are an admixture of elementary and composite states.

This elementary-composite admixture of the massless eigenstate can now be used to explain the fermion mass hierarchy~\cite{Contino:2004vy}, and then predict the sfermion mass spectrum. Consider elementary chiral fermions, $\psi_{L,R}$, that are coupled to the elementary Higgs field, $H$, via the Yukawa interaction $\lambda \, \psi_L^\dagger \psi_R H + h.c.$, where $\lambda$ is an order-one proto-Yukawa coupling (for simplicity, we assume one fermion generation and ignore the distinction between $H_u$ and $H_d$). Diagonalizing the fermion Lagrangian with the Higgs contribution gives the Yukawa coupling expression
\begin{equation}
  y_{\psi}
    \simeq
      \begin{cases}
        \frac{\lambda}{\zeta_\Phi} (\delta - 1) \frac{16 \pi^2}{N} & \phantom{\; 0 \leq} \delta \geq 1 \, , \\[4pt]
        \frac{\lambda}{\zeta_\Phi} (1 - \delta) \frac{16 \pi^2}{N}
        \bigl( \frac{\Lambda_{\text{IR}}}{\Lambda_{\text{UV}}} \bigr)^{2(1-\delta)} & 0 \leq \delta < 1 \, ,
      \end{cases}
\label{eq:fermionmass}
\end{equation}
where we have assumed that $\delta \equiv \delta_L = \delta_R$. We clearly see that when $\delta \geq 1$ (corresponding to a mostly elementary fermion), the Yukawa coupling is of order one for sufficiently large $N$. Conversely, when $0 \leq \delta < 1$ (corresponding to a sizeable composite admixture), the Yukawa coupling has a power-law suppression that depends on the degree of compositeness. This explains why composite fermions (identified with the first- and second-generation Standard Model fermions) have small Yukawa couplings, while elementary fermions (such as the top quark) have order-one Yukawa couplings.

The composite sector is also responsible for supersymmetry breaking. Soft scalar masses are generated
only for the composite sector fields since there is no direct coupling of the supersymmetry breaking to elementary fields. For example, the
massive scalar field, $\phi^{(1)}$ obtains a soft mass
\begin{equation}
  \xi_4\frac{g_\Phi^{(1)2}}{\Lambda_{\text{IR}}^2}
  \left[ \mathcal{X}^\dagger \mathcal{X} \Phi^{(1)\dagger} \Phi^{(1)} \right]_D
    = \xi_4 g_\Phi^{(1)2} \frac{|F_{\mathcal{X}}|^2}{\Lambda_{\text{IR}}^2} \phi^{(1)\dagger} \phi^{(1)} \, ,
\end{equation}
where $\mathcal{X} = \theta \theta F_\mathcal{X}$ is a composite-sector spurion and $\xi_4$ is a dimensionless parameter.
Given the scalar admixture \eqref{eq:scalaradmix}, the corresponding sfermion mass-squared becomes:
\begin{equation}
  \widetilde{m}^2
    \simeq
      \begin{cases}
        \frac{(\delta - 1)}{\zeta_\Phi}
        \frac{16 \pi^2}{N}
        \frac{|F_{\mathcal{X}}|^2}{\Lambda_{\text{IR}}^2}
        \bigl( \frac{\Lambda_{\text{IR}}}{\Lambda_{\text{UV}}} \bigr)^{2(\delta-1)} & \phantom{\; 0 \leq} \delta \geq 1 \, , \\[6pt]
        \frac{(1 - \delta)}{\zeta_\Phi}
        \frac{16 \pi^2}{N}
        \frac{|F_{\mathcal{X}}|^2}{\Lambda_{\text{IR}}^2} & 0 \leq \delta < 1 \, ,
      \end{cases}
   \label{eq:sfermionmass}
\end{equation}
where, for a large-$N$ gauge theory, $\xi_4 \simeq 16 \pi^2 / N$~\cite{Witten:1979kh}.
When the sfermion is mostly elementary ($\delta \geq 1$), the soft mass is power-law suppressed since the supersymmetry breaking is transmitted via the elementary-composite mixing. (Note, however, that for sufficiently 
large $\delta$, radiative corrections will become increasingly important.) This contrasts with the case 
$0 \leq \delta < 1$, where the mass eigenstate is mostly composite and there is no power-law suppression.  Thus, elementary sfermions (identified with the stops) are much lighter than the composite sfermions (identified with the first- and second-generation sfermions), predicting an inverted mass hierarchy!

%%%%%%%%%%%%%%%%%%%%% FIGURE %%%%%%%%%%%%%%%%%%%%%%%%%%%%%%%%%
\begin{figure}[t]
\includegraphics[width=0.4\textwidth]{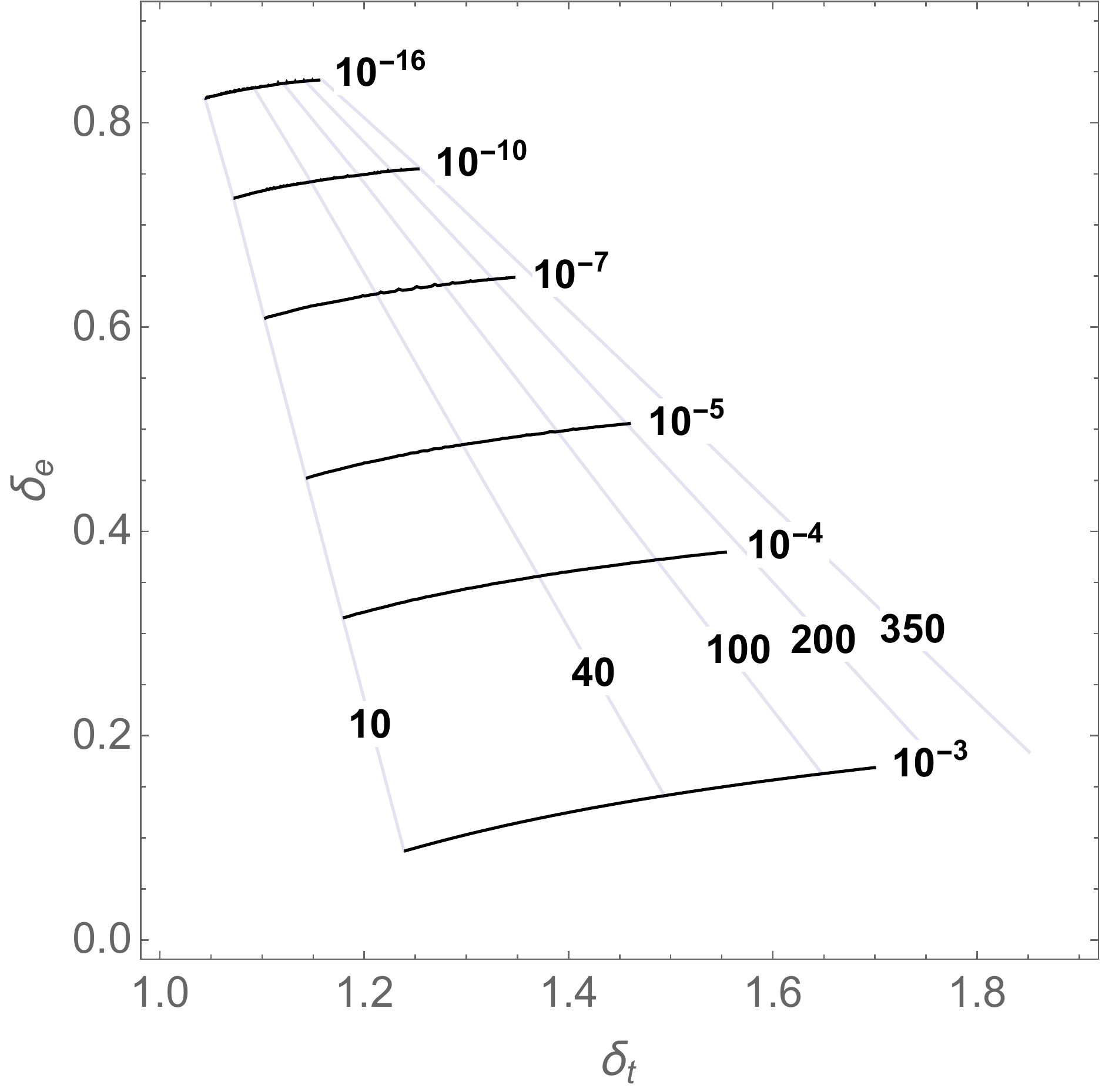}
\caption{The estimated range of anomalous dimensions $\delta_e, \delta_t$ that gives rise to the observed
hierarchy between the electron and top quark Yukawa couplings, assuming $\Lambda_{\text{UV}} = 10^{18}$~GeV, $\tan \beta = 3$, and a soft mass threshold at 50~TeV. The slanted horizontal and vertical lines are contours of the ratio $\Lambda_{\text{IR}}/\Lambda_{\text{UV}}$ and the sfermion mass ratio $m_{\widetilde{e}}/m_{\widetilde{t}}$, respectively. }
\label{fig:fermionratio}
\end{figure}
%%%%%%%%%%%%%%%%%%%%%%%%%%%%%%%%%%%%%%%%%%%%%%%%%

The fermion and sfermion mass hierarchies critically depend on the anomalous dimensions $\delta$.
To illustrate this, we consider the $\delta_{e,t}$ values required to obtain the electron top-quark Yukawa
coupling ratio $y_e/y_t$ at the IR scale. These are plotted in Fig.~\ref{fig:fermionratio} for various values of
$\Lambda_{\text{IR}}/\Lambda_{\text{UV}}$, where the ratio $y_e/y_t$ (${\sim}10^{-5}$) at the IR scale is determined via two-loop renormalization group evolution (assuming a universal soft mass threshold).
Using \eqref{eq:sfermionmass}, this then predicts the sfermion mass ratio $m_{\widetilde{e}}/m_{\widetilde{t}}$ at the IR scale. As shown in Fig.~\ref{fig:fermionratio}, the allowed region is $0 \lesssim \delta_e \lesssim 0.9$ and $1 \lesssim \delta_t \lesssim 1.8$, depending on the value of $\Lambda_{\text{IR}}/\Lambda_{\text{UV}}$.
The largest value of the ratio $m_{\widetilde{e}}/m_{\widetilde{t}}$ is approximately $140~(390)$ for  $\Lambda_{\text{IR}}/\Lambda_{\text{UV}} \simeq 10^{-3}~(10^{-16})$.
Note that the slanted horizontal contours in Fig.~\ref{fig:fermionratio} end on the right, at the $\delta_t$ value
for which radiative corrections to the soft mass \eqref{eq:sfermionmass} begin to dominate. These corrections are calculated in Ref.~\cite{BGM}.

A partially-composite analysis can also be done for the vector and gravity supermultiplets. They lead to a mostly elementary gauge boson and gaugino, and an elementary graviton and gravitino~\cite{Batell:2007jv}. Since supersymmetry breaking occurs in the composite sector, this implies that the gauginos are lighter than the mostly composite first- and second-generation sfermions and comparable in mass to the mostly elementary third-generation sfermions. On the other hand, since the gravitino has a tiny composite admixture, it is almost always the lightest supersymmetric particle (LSP). These are the qualitative features of the partially-composite sparticle spectrum. Further details are presented in Ref.~\cite{BGM}.

\vspace{-3mm}
\section{A 5D gravity model}
\label{sec:5Dmodel}
\vspace{-3mm}

The partially-composite supersymmetric framework generically relates the fermion and
sfermion mass spectra that result from some (unknown) strong dynamics. In order to model the underlying dynamics and obtain quantitative predictions, we now consider a five-dimensional (5D) dual gravity model that is motivated by the AdS/CFT correspondence~\cite{Maldacena:1997re}. The 5D spacetime, $(x^\mu, y)$, where $\mu = 0, 1, 2, 3$ labels the four-dimensional (4D) coordinates and the fifth coordinate, $y$, is compactified on an orbifold ($S^1/\mathbb{Z}_2$). The anti-de Sitter (AdS) metric is given by
\begin{equation}
\label{eq:AdS5metric}
  ds^2 = e^{-2 k y} dx^2 + dy^2 \, ,
\end{equation}
where $k$ is the AdS curvature scale.
The 5D spacetime is a slice of AdS$_5$ bounded by two 3-branes located at the orbifold fixed points: a UV brane at $y = 0$ and an IR brane at $y = \pi R$, where $R$ is the orbifold radius~\cite{Randall:1999ee}.

Besides gravity, we introduce the full matter and gauge-sector content of the minimal supersymmetric standard 
model in the AdS$_5$ bulk. The $\mathcal{N} = 1$ chiral matter and vector superfields are embedded into 5D $\mathcal{N} = 1$ hypermultiplets and vector supermultiplets, respectively. The 4D superfields are then identified with the massless zero modes, while the massive Kaluza-Klein states form 4D $\mathcal{N} = 2$ supermultiplets with masses of order $\Lambda_{\text{IR}}$. The $\mathcal{N} = 1$ Higgs supermultiplets, meanwhile, are 4D fields confined to the UV brane. In this setup, each fermion zero mode obtains a mass from a UV boundary-bulk Yukawa interaction with 5D Yukawa coupling, $Y^{(5)}$. The fermion mass hierarchy then arises from the overlap of the UV-localized Higgs fields with the left- and right-handed bulk fermion fields with profiles $\psi_{L,R} \propto e^{(\frac{1}{2}\mp c) k y}$ in the fifth dimension, where the $c$ are dimensionless bulk fermion mass parameters~\cite{Grossman:1999ra, Gherghetta:2000qt}. Once the $c$ parameters are determined for each fermion flavor, the sparticle mass spectrum can then be predicted.

Supersymmetry is only broken on the IR brane and can be parameterized by introducing a boundary interaction with a spurion $X = \theta^2 F_X$ for each 5D hypermultiplet $\Phi(x^\mu,y)$:
\begin{equation}
 \int d^5 x \, \sqrt{-g} \int d^4 \theta \, \frac{X^\dag X}{\Lambda_{\text{UV}}^2 k} \, \Phi^\dag \Phi \, \delta(y - \pi R) \, .
\label{eq:sfermionspurion}
\end{equation}
This interaction leads to the sfermion soft mass
\begin{equation}
  \widetilde{m}_{L,R}
    \simeq
      \begin{cases}
             \left(\pm c-\frac{1}{2}\right)^{1/2} \frac{F}{\Lambda_{\text{IR}}} \, e^{(\frac{1}{2} \mp c)\pi k R} & \pm  c > \frac{1}{2} \, , \\[4pt]
             \left(\frac{1}{2}\mp c \right)^{1/2} \frac{F}{\Lambda_{\text{IR}}} & \pm c < \frac{1}{2} \, ,
      \end{cases}
      \label{eq:KKscalar}
\end{equation}
where $\sqrt{F}\equiv \sqrt{F_X} e^{-\pi k R}$, and the back-reaction on the sfermion profile by the boundary mass
is negligible (\textit{i.e.}, $\sqrt{F}/\Lambda_{\text{IR}}\lesssim 1$). Furthermore, note that possible flavor off-diagonal mass terms in (\ref{eq:sfermionspurion}) have been neglected since the sfermion mass scale is assumed to be ${\cal O}(100)$\,TeV. Using the AdS/CFT dictionary relations $\Lambda_{\text{IR}}/\Lambda_{\text{UV}} =e^{-\pi kR}$ and $\delta = |c \pm \frac{1}{2}|$, the expressions \eqref{eq:KKscalar} are seen to be consistent with the masses \eqref{eq:sfermionmass} obtained in the 4D holographic theory. Quantum corrections to the tree-level scalar masses \eqref{eq:KKscalar} arising from loops of bulk hypermultiplets and vector supermultiplets are important for suppressed masses ($\pm c \gtrsim \frac{1}{2}$). These are computed in Ref.~\cite{BGM}, and their effect is typically to reduce the sfermion mass hierarchy.

%%%%%%%%%%%%%%%%%%%%%%%%%%%%%%%%%%%%%%%%%%%%%%%%%%%%%%%%%%%%
\begin{figure*}[t]
  \centering
  \includegraphics{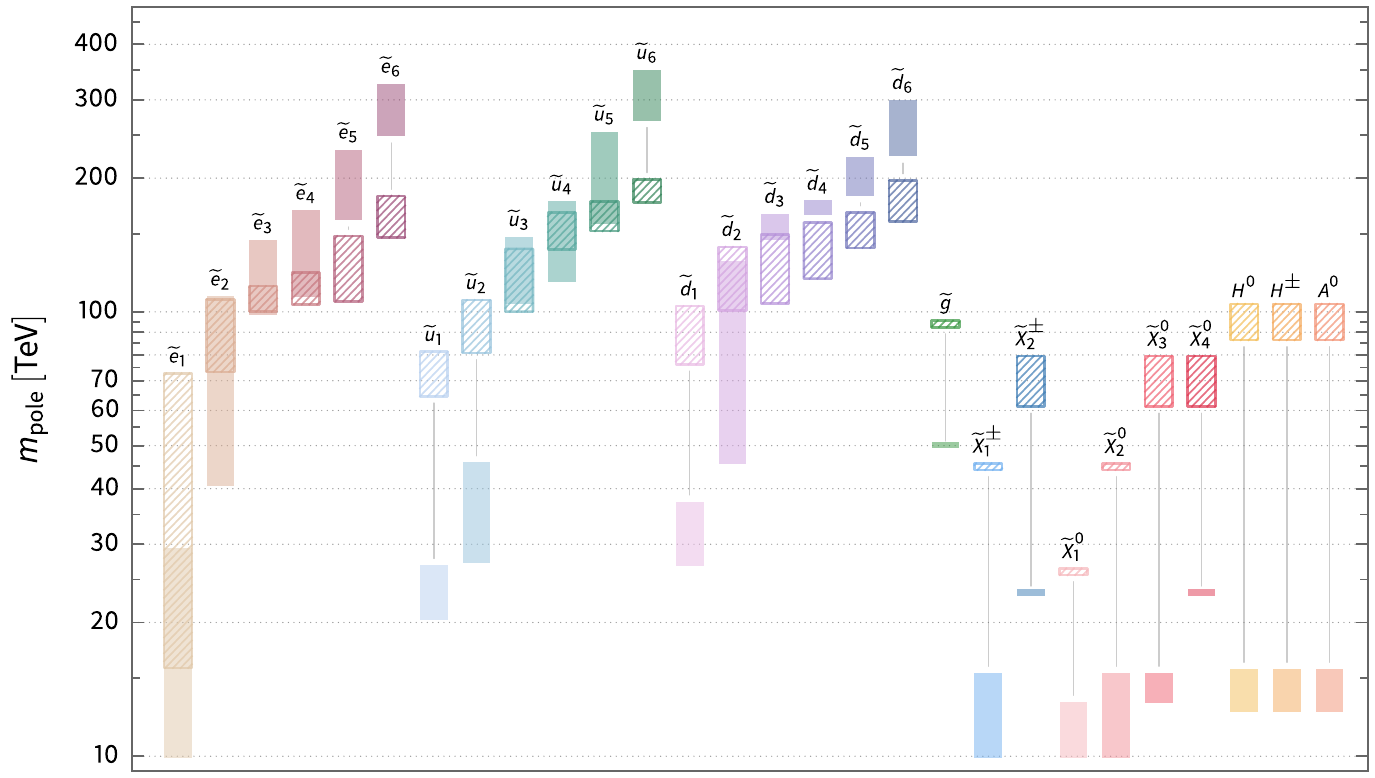}
  \caption{The sparticle mass spectrum for two benchmark scenarios: singlet spurion case (hatched) with $\Lambda_{\text{IR}} = 2 \times 10^{16}$~GeV, $\sqrt{F} = 4.75 \times 10^{10}$~GeV, $\tan \beta \sim 3$ and $Y^{(5)} k = 1$; and non-singlet spurion case (solid) with $\Lambda_{\text{IR}} = 6.5 \times 10^{6}$~GeV, $\sqrt{F} = 2 \times 10^{6}$~GeV, $\tan \beta \sim 5$ and $Y^{(5)} k = 1$.}
  \label{fig:running}
\end{figure*}
%%%%%%%%%%%%%%%%%%%%%%%%%%%%%%%%%%%%%%%%%%%%%%%%%%%%%%%%%%%%

Similarly, introducing an IR-boundary gaugino interaction term $X W^a_\alpha W^{a\alpha}$, where $W^a_\alpha$ is the gauge field strength superfield, gives rise to gaugino masses $M_{\lambda_a} \simeq g_a^2 F/\Lambda_{\text{IR}}$~\cite{Chacko:2003tf}, with $g_a\,(a=1,2,3)$ the corresponding Standard Model gauge couplings.
The gaugino masses are suppressed relative to the heavier sfermions (with $\pm c < \frac{1}{2}$).
Alternatively, if the supersymmetry-breaking sector does not contain any gauge singlets, the gaugino masses may instead be generated by a boundary interaction $X^\dagger X W^a_\alpha W^{a\alpha}$. This leads to 
gaugino masses, $M_{\lambda_a} \simeq g_a^2 F^2/\Lambda^3_{\text{IR}}$ that are further suppressed.

When supersymmetry is spontaneously broken on the IR boundary, the effective 4D cosmological constant receives a positive contribution from $F_X$. In the 5D warped geometry, this contribution can be canceled by the addition of a constant superpotential $W$ on the UV brane \cite{Randall:1998uk, Luty:2000ec, Luty:2002ff, Gherghetta:2002nr, Chacko:2003tf, Itoh:2006fv, Gherghetta:2011wc}, giving rise to a gravitino mass $m_{3/2}\simeq F/(\sqrt{3} M_P)$. Since the gravitational coupling is Planck-scale suppressed, the gravitino mass is lower than the characteristic soft-mass scale $F/\Lambda_{\text{IR}}$ by a warp factor.

The Higgs sector does not couple directly to the IR brane, and therefore the Higgs soft terms $m_{H_u}^2$, $m_{H_d}^2$, and $b$
as well as the trilinear soft scalar couplings ($a$-terms)
at the IR-brane scale are zero at tree level.
However, these soft terms are generated via radiative corrections from their interactions with bulk hypermultiplets and vector supermultiplets~\cite{BGM}.
The resulting values for the Higgs soft masses, obtained at the IR-brane scale, must be run down to near the electroweak scale in order to check that electroweak symmetry is broken.
The Higgs $\mu$-term is assumed to arise on the UV brane from a higher-dimensional superpotential term allowed by an extra $U(1)$ symmetry, as in the Kim-Nilles mechanism~\cite{Kim:1983dt}. Its value, along with $\tan \beta$ (the ratio of the Higgs vacuum expectation values), is determined by the conditions for electroweak symmetry breaking.

The parameters of the 5D model therefore consist of the IR brane scale, $\Lambda_{\text{IR}}$, and the supersymmetry breaking scale, $\sqrt{F}$. In addition, there is a universal 5D Yukawa coupling, $Y^{(5)}$,
and the six bulk fermion mass parameters $c_{L_i,Q_i}$ (one for each generation of leptons and quarks).
These parameters can be used to determine both the fermion and sparticle mass spectra. However, there are a number of phenomenological and theoretical constraints which restrict the possible parameter values. These include:

\begin{itemize}[itemsep=0pt,leftmargin=1em,partopsep=0pt]

\item \textbf{Gravitino dark matter:} Assuming $R$-parity conservation, the gravitino LSP makes an excellent dark matter candidate, provided $m_{3/2} \gtrsim 1$~keV~\cite{Bolz:2000fu,Pradler:2006hh}.

\item \textbf{Higgs mass and electroweak symmetry breaking:} The observed 125~GeV Higgs boson can be accommodated if the mass of the lightest stop is $\mathcal{O}(10)$~TeV. Since the Higgs-sector soft terms are generated radiatively, the requirement that the Higgs scalar potential correctly breaks electroweak symmetry leads to further indirect constraints on the soft masses of the sfermions.

\item \textbf{Supersymmetric flavor problem:} To avoid generating excessive flavor-changing processes, the first- and second-generation sfermions must be at least 100~TeV.

\item \textbf{Gauge coupling unification:} To preserve the successful supersymmetric prediction of gauge coupling unification (assuming any underlying dynamics is SU(5) symmetric), the gaugino and Higgsino masses must be lighter than 300~TeV.

\item \textbf{Charge- and color-breaking minima:} Since the predicted sfermion mass spectrum at $\Lambda_{\text{IR}}$ is
flavor-dependent and the first- and second-generation sfermions are typically hierarchically larger than the third-generation sfermions, there are both one-loop $D$-term and two-loop gauge contributions to scalar masses that can lead to charge and color-breaking minima.

\end{itemize}

Subject to the above constraints, we choose two benchmark scenarios corresponding to the singlet and non-singlet spurion cases. The singlet case has parameter values $\Lambda_{\text{IR}} = 2 \times 10^{16}$~GeV, $\sqrt{F} = 4.75 \times 10^{10}$~GeV, $Y^{(5)} k = 1$, and $\tan\beta \simeq 3$ at the
IR-brane scale, whereas the non-singlet case has parameter values  $\Lambda_{\text{IR}} = 6.5 \times 10^{6}$~GeV, $\sqrt{F} = 2 \times 10^{6}$~GeV, $Y^{(5)} k = 1$, and $\tan\beta \simeq 5$ at the
IR-brane scale. The sfermion pole mass predictions are presented in Fig.~\ref{fig:running}, where the spread in the masses results from a scan over the $c$-parameters in order to fit the Yukawa coupling hierarchy. The Higgs mass lies in the range 124--126~GeV, with $\operatorname{sign} \mu = -1$ . Furthermore, the mass of the LSP gravitino is 535~GeV (1~keV) for the singlet (non-singlet) spurion case. The sfermion masses obtained directly result from explaining the fermion mass hierarchy. They reveal a distinctive, flavor-dependent inverted mass hierarchy, in contrast to usual supersymmetric models where scalar and gaugino masses are unconstrained by the fermion mass spectrum.

\vspace{-3mm}
\section{Conclusion}
\label{sec:conclusions}
\vspace{-3mm}

In this Letter, we have presented a partially-composite supersymmetric model that assumes the first two generations of matter are (mostly) composite, while the Higgs and third generation matter are (mostly) elementary. This feature can then be used to explain the fermion mass hierarchy, predicting, as a consequence, a
distinct sparticle mass spectrum with an inverted sfermion mass hierarchy: light stops and staus and heavy first-and second-generation sfermions. The underlying dynamics responsible for the compositeness can be modelled with a dual 5D gravity theory that further predicts a gravitino LSP, together with gauginos and Higgsinos ranging from the lightest neutralino at 10~TeV to gluinos at 90~TeV. These masses are split from the heavier first- and second-generation sfermions, thereby preserving the successful supersymmetric prediction of gauge coupling unification.
A more detailed analysis of this model is given in Ref.~\cite{BGM}.

The partially-composite supersymmetric model intricately connects the generation of the fermion
mass hierarchy with the sfermion masses. It is striking that the predicted sparticle spectrum seems
to provide an appealing fit to the current experimental constraints. While not directly accessible
at the 13~TeV LHC, the signatures of this sparticle spectrum, such as distinctive long-lived NLSP decays, may be within the reach of a future high-energy collider. Alternatively, the heavy first- and second-generation sfermions could be indirectly probed at flavor-violation experiments such as the Mu2e experiment~\cite{Bartoszek:2014mya} or at experiments aiming to measure the electric dipole moment of the electron~\cite{Andreev:2018ayy}. Thus, partial compositeness and supersymmetry are intriguing possibilities that 
could together play a central role in addressing some of the shortcomings of the Standard Model.

\vspace{0.5cm}
\noindent
{\bf Acknowledgments}
We thank Jason Evans, Ben Harling, and Alex Pomarol for helpful discussions. This work is supported in part by
the U.S. Department of Energy under Grant DE-SC0011842 at the University of Minnesota.

\end{document}